\documentclass[useAMS,usenatbib]{mnras}
\usepackage{graphicx}
\usepackage{amsmath,amssymb}
\usepackage{epstopdf}
\usepackage{bm} 
\newcommand{\del}{\partial}

\newcommand{\ALP}{{\boldsymbol{\alpha}}}
\newcommand{\largec}{{\boldsymbol{C}}}

\newcommand{\x}{{\boldsymbol{x}}}
\newcommand{\y}{{\boldsymbol{y}}}

\newcommand{\f}{\frac}

\newcommand{\BF}{\begin{figure}\begin{center}}
\newcommand{\EF}{\end{center}\end{figure}}
\newcommand{\BE}{\begin{equation}}
\newcommand{\EE}{\end{equation}}
\newcommand{\BEA}{\begin{eqnarray}}
\newcommand{\EEA}{\end{eqnarray}}
\newcommand{\ti}{\textit}
\newcommand{\tr}{\textrm}

\def\v#1{\boldsymbol #1}

\newcommand{\ms}{M_{\odot}}

\begin{document}
\title[On the Origin of Flux Ratio Anomaly in Quadruple Lens Systems]{
On the Origin of Flux Ratio Anomaly in Quadruple Lens Systems}
\author[Kaiki Taro Inoue]
{Kaiki Taro Inoue$^1$\thanks{E-mail:kinoue@phys.kindai.ac.jp}
\\
$^{1}$Faculty of Science and Engineering, 
Kindai University, Higashi-Osaka, 577-8502, Japan  }
\date{\today}

\pagerange{\pageref{firstpage}--\pageref{lastpage}} \pubyear{0000}
\maketitle
\label{firstpage}
\begin{abstract}
We explore the origin of flux ratio anomaly in quadruple lens systems. 
Using a semi-analytic method based on 
$N$-body simulations, we estimate the effect of possible magnification 
perturbation caused by subhaloes with a mass scale of $ \lesssim 10^9\,h^{-1} \ms$ in lensing
 galaxy haloes. Taking into account astrometric shifts, assuming that the primary lens is
 described by a singular isothermal ellipsoid, the expected change to the flux ratios per a multiply 
lensed image is just a few percent and the mean of the expected convergence
 perturbation at the effective Einstein radius of the 
lensing galaxy halo is $\langle \delta \kappa_{\tr{sub}} \rangle  = 0.003$, corresponding to
the mean of the ratio of a projected
dark matter mass fraction in subhaloes 
at the effective Einstein radius $\langle f_{\tr{sub}} \rangle  = 0.006$.
In contrast, the expected change to the flux ratio
caused by line-of-sight structures is typically $\sim 10$ percent and
 the mean of the convergence perturbation is 
$\langle |\delta \kappa_{\tr{los}}| \rangle = 0.008$, corresponding to $\langle
 f_{\tr{los}} \rangle  = 0.017$. 
The contribution of magnification perturbation caused
 by subhaloes is $\sim 40$ percent of the total at a source redshift
 $z_\tr{S}= 0.7$ and decreases monotonically in $z_\tr{S}$ to $\sim
20$ percent at $z_\tr{S}= 3.6$. Assuming statistical isotropy, the convergence
 perturbation estimated from observed 11 quadruple lens systems has 
a positive correlation with the source redshift $z_\tr{S}$, which is much stronger
than that with the lens redshift $z_{\tr{L}}$. This feature also supports an idea
 that the flux ratio anomaly is caused mainly by
line-of-sight structures rather than subhaloes. We also discuss about
a possible imprint of line-of-sight structures in demagnification 
of minimum images due to locally underdense structures in the line of sight. 
\end{abstract}

\section{Introduction}
It has been known that some quadruply lensed quasars show anomalies in the observed 
flux ratios of lensed images provided that 
the gravitational potential of the lensing galaxy halo is
sufficiently smooth. Such a discrepancy is 
called the ``anomalous flux ratio'' and 
has been considered as an imprint of cold dark matter (CDM) subhaloes
with a mass of $\sim 10^{8-9} \ms$ in the parent
galaxy haloes \citep{mao1998,metcalf2001,chiba2002,dalal-kochanek2002,
keeton2003, inoue-chiba2003, kochanek2004, metcalf2004,chiba2005,sugai2007,mckean2007,more2009,
minezaki2009, xu2009,xu2010, fadely2012, macleod2013}. 

However, intergalactic haloes in the line of sight 
can act as perturbers as well \citep{chen2003,metcalf2005a,xu2012}.
Indeed, taking into account astrometric shifts, recent studies 
have found that the observed anomalous flux ratios can be 
explained solely by line-of-sight structures with a surface density $\sim 
10^{7-8}\, h^{-1}\ms/\textrm{arcsec}^2$ \citep{inoue-takahashi2012,
takahashi-inoue2014,inoue-etal2015, inoue-minezaki2015} without taking into account 
subhaloes in the lensing galaxies. Moreover, using the cusp and fold
caustic relations expressed in statistics 
$R_{\tr{cusp}}$ and $R_{\tr{fold}}$, \citet{xu2015}
found that the perturbation caused by subhaloes is not sufficient for
explaining the observed anomalies. 

In order to model quadruple lens systems, one must take into account 
other massive objects that may significantly affect the flux ratios 
\citep{mckean2007, more2009}. Moreover, disk components can also significantly perturb
the flux ratios \citep{hsueh2016}. Thus, it is sometimes difficult to make
constraints using the cusp and fold caustic relations because they
hold only in lenses with a smooth potential and small angles between
two bright lensed images. 

In this paper, we investigate the origin of flux ratio anomaly
in 11 quadruple lens systems including radio and
mid-infrared (MIR) fluxes. We do not use the ``traditional'' statistics
$R_{\tr{cusp}}$ and $R_{\tr{fold}}$ in the literature. The reasons are: 1)
the lens systems can have more complex structures such as a secondary
and a third lens. 2) the cusp and fold caustic relations do not exactly
hold because the angle between two images in a triplet or a double is
not necessarily small. 3) the magnification
bias in the selection of lens candidates can significantly affect the
statistics. We also use the information of possible astrometric shifts,
which was often ignored in the literature (see, however, \citet{chen2007,
sluse2012}). Furthermore, we pay attention to the correlation
between the perturbation to the fluxes and the redshifts of the primary
lenses and the source, which would shed a new light on the origin of
the flux ratio anomaly.

In section 2, we estimate the perturbation effects caused by subhaloes in 
lensing galaxies using a semi-analytic mass function of subhaloes. In
section 3, we briefly describe our method for calculating 
the perturbation effects caused by line-of-sight structures. In section
4, we briefly describe the 11 quadruple lenses. In section 5, we explain
the adopted lens model of an unperturbed system. In section 6, we explain our
method using a statistic ``$\eta$'' for fitting the model to the data. In section 7, we show our
result for the magnification and convergence perturbations and
their correlations with the lens and source redshifts.
In section 8, we briefly discuss about a new type of anomaly that can be
used for distinguishing the origin of the perturbation.

In what follows, we assume a cosmology 
with a matter density $\Omega_{m,0}=0.3134$, a baryon density 
$\Omega_{b,0}=0.0487$, a cosmological constant $\Omega_{\Lambda,0}=0.6866$,
a Hubble constant $H_0=67.3\, \textrm{km}/\textrm{s}/\textrm{Mpc}$,
a spectral index $n_s=0.9603$, and the root-mean-square (rms) 
amplitude of matter fluctuations at $8 h^{-1}\, \textrm{Mpc}$, 
$\sigma_8=0.8421$, which are obtained from the observed 
CMB (Planck+WMAP polarization; \citet{ade2014}). $m$ and $r$ refer to the mass and radius of
a subhalo, and $M$ and $R$ refer to the mass of the host and
the radial location within a host. The virial mass $M_{200}$ is 
the mass inside the virial radius $R_{200}$ inside
which the average matter density equals 200 times the critical density
of the universe.

\section{Substructure Lensing}
In order to estimate the magnification perturbation due to 
subhaloes in a lensing galaxy halo, we need to calculate the mass 
function of subhaloes. 

At a proper distance $R$ from the centre of a 
host halo, the number density of subhaloes $n$ per a logarithmic interval of
the mass $m_{in}$ of subhaloes at the infall is well approximated by
\BE
\f{dn(m_{in},R)}{d  \ln m_{in}} \propto m_{in}^{-\alpha} \rho(R),
\EE
where $\rho(R)$ is the density of the host halo at $R$ and
$\alpha \sim 0.9$\citep{vandenbosch2005}\footnote{In the literature, the other definition 
$\tilde{\alpha}=\alpha+1$ is sometimes used.}. 

However, after infalling, the masses of subhaloes are
stripped off by the tidal force of the host halo, and some of them are
completely disrupted. Therefore, the mass $m$ of a subhalo
at the time of deflection of photons can be significantly 
reduced at the position of lensed images. 

Assuming that subhaloes have a density profile similar to that of an 
singular isothermal sphere (SIS), the bound fraction $\mu=m/m_{in}$ of subhaloes
at $R$ can be approximated as
\BE
\mu(R) \propto R^\beta,
\EE   
where $\beta \sim 1$ if the kinetic energy of subhaloes is negligible. 
In real setting, one must also 
consider the kinetic energy of subhaloes at the time of
infall and the change of tidal stripping that depend both on the orbit and
the background gravitational field. If tidal stripping of
each subhalo can be treated as an uncorrelated random process, one may approximate
the set of bound fraction $\mu$ for each subhalo as random variables.  
Assuming that $\mu$ is written in terms of a polynomial and 
obeys a log-normal distribution and the the spatial distribution with a
given infall mass traces the
mass density profile, based on two sets of high-resolution 
cosmological $N$-body simulations; the Aquarius \citep{springel2008}
and Phoenix \citep{gao2012} simulation suites, \citet{han2015} 
obtained a subhalo mass
function for a host halo with $R_{200}$ and $M_{200}$,
\BE
\f{dn(m,R)}{d  \ln m} =A_{acc}B f_s e^{\sigma^2 \alpha^2/2}
 \biggl[\f{m}{m_0 \bar{\mu}(R) }  \biggr]^{-\alpha} \f{\rho(R)}{m_0},
\label{eq3}
\EE 
where the mean bound fraction is 
\BE
\bar{\mu}(R)=\mu_* \biggl(\f{R}{R_{200}} \biggr)^\beta
\EE   
and 
\BEA
\mu_*&=&0.5(M_{200}/m_0)^{-0.03},~~\alpha=0.95,
\nonumber
\\
\beta&=&1.7(M_{200}/m_0)^{-0.04},~~ A_{acc}=0.1(M_{200}/m_0)^{-0.02} 
\nonumber
\\
f_s&=&0.55,~~ m_0=10^{10} h^{-1}\ms.
\EEA
Let us assume that a parent halo that hosts a primary lensing galaxy 
has an NFW profile with a concentration
parameter $C$ and a critical density $\rho_{crit}$ at a redshift $z$,  
\BE
\rho^{NFW}(R)=\f{\delta_c(C)\rho_{crit}(z)}{(R/R_{-2})(1+R/R_{-2})^2},
\EE
where 
\BE
R_{-2}=R_{200}/C, ~~~~ \delta_c(C)=\f{200}{3}\f{C^3}{\ln[1+C]-C/(1+C)}.
\EE
Integrating equation (\ref{eq3}) over the line of sight, we obtain the
mass function projected onto the lens plane as a function of $m$ and 
the projected proper distance $R_{2D}$ to the host halo center in the
lens plane as
\BEA
\f{d n_s(m,R_{2D})}{d  \ln m}& =& A_{acc}B f_s e^{\sigma^2 \alpha^2/2}
 \biggl[\f{m}{m_0}  \biggr]^{-\alpha}
 \f{\delta_c(C)\rho_{crit}(z)}{m_0}
\nonumber
\\
&\times& K[R_{2D},R_{200}/C,\gamma],
\label{eq8}
\EEA
where $n_s$ is the surface number density, $\gamma=\alpha\beta$ and 
\BEA
\lefteqn{K[R_{2D},R_{200}/C,\gamma]}
\nonumber
\\
&=&
2^{2-\gamma}\mu_*^\alpha 
\f{R_{200}^{-\gamma}R_{2D}^{-3+\gamma}
R_{-2}^3}{\Gamma[3-\alpha \beta] (R_{2D}^2-R_{-2}^2)}
\nonumber
\\
&\times& 
\Biggl[ (\Gamma[3/2-\gamma/2] )^2 R_{-2} 
\biggl((-2+\gamma)R_{2D}^2-\bigl(R_{-2}^2+\gamma(R_{2D}^2
\nonumber
\\
&-&R_{-2}^2)\bigr) 
~{_2F_1}\biggl[1,\f{3-\gamma}{2},2-\f{\gamma}{2},\f
{R_{-2}^2}{R_{2D}^2}\biggr] \biggr) 
\nonumber
\\
&+&
\Gamma[1-\gamma/2]\Gamma[2-\gamma/2]
R_{2D}(R_{2D}^2-R_{-2}^2)
\nonumber
\\
&\times&{_3F_2}\biggl[1,\f{3}{2},\f{1-\gamma}{2};
\f{1}{2},\f{3-\gamma}{2};\f{R_{-2}^2}{R_{2D}^2}\biggr]
\Biggr].
\label{eq9}
\EEA
Equations (\ref{eq8}) and (\ref{eq9}) give the surface mass density $\Sigma_s$
and the surface number density $n_s$ of subhaloes at a distance $R_{2D}$
in an NFW halo,
\BEA
\Sigma_s(R_{2D})&=&\int_{m_{min}}^{m_{max}} \f{d n_s(m,R_{2D})}{d  \ln m} dm,
\\
n_s(R_{2D})&=&\int_{m_{min}}^{m_{max}} \f{d n_s(m,R_{2D})}{d  \ln m} d
\ln m,
\EEA
where $m_{min}$ and $m_{max}$ are the minimum and the maximum mass
of the subhaloes. 
\vspace*{0.5cm}
\begin{figure*}
\includegraphics[width=170mm]{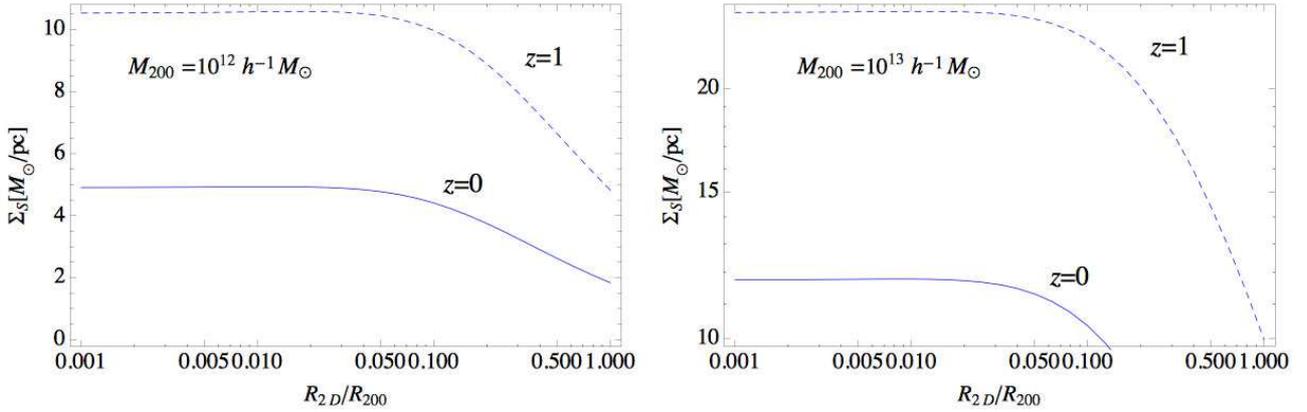}
\caption{Surface mass density of subhaloes at a normalised
 distance $R_{2D}/R_{200}$ for $M_{200}=10^{12}~h^{-1}\ms $ (left) and $M_{200}=10^{13}~h^{-1}\ms $  (right). }
\label{sigma}
\end{figure*}
\begin{figure*}
\includegraphics[width=170mm]{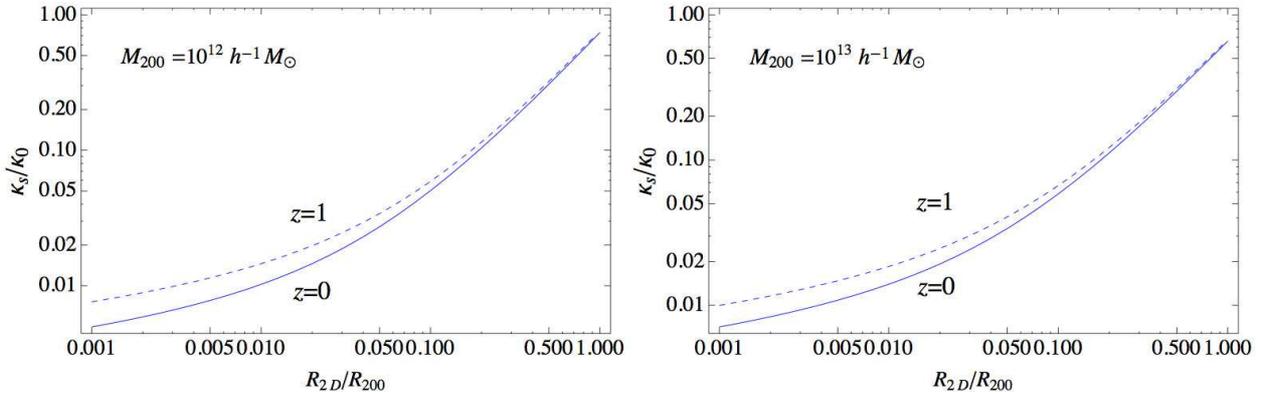}
\caption{Ratio of the convergence of subhaloes to that of the host NFW halo at a normalised
 distance $R_{2D}/R_{200}$ for $M_{200}=10^{12}~h^{-1}\ms $ (left) and $M_{200}=10^{13}~h^{-1}\ms $  (right). }
\label{kappa-ratio}
\end{figure*}
Note that the obtained subhalo surface number densities are consistent
with those in \citep{xu2015} in which 500 random projections are 
used per halo in the Aquarius and Phoenix simulations.
As we shall show in section 7, the effective Einstein radius is 
$R_E \sim 0.01R_{200}$. As shown in Fig. \ref{sigma}, at projected
proper distances $R_{2D} \lesssim 0.03 R_{200}$, for
$M_{200}=10^{12-13}h^{-1}\ms $, the surface mass density
is nearly constant due to tidal disruption.
Since for $R_{2D}<R_{-2}$, most of the surface mass density of an NFW (or SIS) halo
is determined by masses whose mass density has a slope of $\sim -2$, the contribution of 
masses at inner regions whose mass density has a slope of $\sim -1$ 
is small. Then the surface mass density can be approximately
written as
\BE
\Sigma_s(R_{2D},m) \sim m^{-\alpha} R_{2D}^{\alpha \beta} R_{2D}^{-1}.
\label{eq12}
\EE
Assuming that $\alpha\sim \beta \sim 1$ for NFW haloes, equation (\ref{eq12})
gives $\Sigma_s(R_{2D},m) \approx m^{-\alpha}$. Thus $\Sigma_s$ does not
depend on $R_{2D}$. Setting $R_{2D}=0$, we can obtain an approximated 
formula for the surface mass density at $R_{2D} \lesssim R_E$,
\BEA
\Sigma_s(R_E) &\approx& 
\f{\mu_*^\alpha m_0^{\alpha-1} }{1-\alpha}
\biggl[m_{max}^{1-\alpha}-m_{min}^{1-\alpha}
 \biggr]
\nonumber
\\
&\times& e^{\sigma^2 \alpha^2/2} J[\gamma,R_{200},C;z],
\EEA
and 
\BEA
n_s(R_E)  &\approx& 
\f{\mu_*^\alpha m_0^{\alpha-1} }{\alpha}
\biggl[-m_{max}^{-\alpha}+m_{min}^{-\alpha}
 \biggr]
\nonumber
\\
&\times& e^{\sigma^2 \alpha^2/2} J[\gamma,R_{200},C;z],
\EEA
where
\BEA
J[\gamma,R_{200},C,z]&\equiv &2 \pi A_{acc}B f_s (1-\gamma)\csc(\gamma
\pi ) R_{200}
\nonumber
\\
&\times&
C^{-(1+\gamma)} \delta_c(C)\rho_{crit}(z).
\EEA
Here a condition $0<\gamma <2$ should be satisfied in order to avoid divergence.

The surface mass density of an NFW lensing galaxy halo is given by
\BEA
\Sigma_0(R_{2D})&=& \Biggl[\f{2 R_{-2}^3}{R_{2D}^2-R_{-2}^2}+\f{4
R_{-2}^4}{(R_{-2}^2-R_{2D}^2)^{3/2}}
\nonumber
\\
&\times&
\textrm{arccoth}\biggl[\f{R_{2D}+R_{-2}}{(R_{-2}^2-R_{2D}^2)^{1/2}}
\biggr]\Biggr]\delta_c(C)\rho_{crit}(z).
\nonumber
\\
\EEA
Then, the ratio of the convergence of subhaloes $\kappa_s$ to that 
of the host NFW halo $\kappa_0$ can be obtained from equations (3)-(11) and (16).  
As shown in Fig. \ref{kappa-ratio}, the ratio increases in $R_{2D}$. 
At $R_{2D}/R_{200}\sim 0.01$, the dependence on $R_{2D}$ is determined solely by
the surface mass density of an NFW host halo as $\kappa_s$ is nearly
constant. The slope is slightly flatter than $R_{2D}^{-1}$, because the
effective Einstein radius is
smaller than the scale radius $R_{2D}/R_{200}\sim 0.01 < R_{-2}/R_{200}
\sim 0.1$. As we shall show later, the ratio $\kappa_s/\kappa_0$ gives
an upper limit for the dark matter fraction $f$ at the position of 
lensed images, which characterizes the strength of contribution from
subhaloes. 
 
\section{Line-of-sight lensing }

To take into account the non-linear effects of line-of-sight structures, 
we first calculate the non-linear power spectra of matter fluctuations down to 
mass scales of $\sim 10^5\, h^{-1 } \ms$ using $N$-body simulations \citep{inoue-takahashi2012,takahashi-inoue2014}. For simplicity, we do not consider baryonic effects 
in our simulations. In the following, we assume that the
astrometric shifts of lensed images caused by perturbers are sufficiently smaller than
the deflection angle of a background lens model and the strong lensing
effects caused by perturbers are negligible.
Then the magnification contrast $\delta^{\tilde{\mu}}=\delta
\tilde{\mu}/ \tilde{\mu} $ for each lensed image 
with a nonperturbed convergence $\kappa$ and a shear $\gamma$
can be approximated as 
\BE
\delta^{\tilde{\mu}}\approx \frac{2(1-\kappa)\delta \kappa + 2\gamma
\delta \gamma  }{(1-\kappa)^2- \gamma^2},
\label{eq:delta}
\EE
where $\delta \kappa$ and $\delta \gamma$ are
the convergence and shear due to perturbers in the line of sight 
\citep{inoue-takahashi2012,takahashi-inoue2014}.
Using equation (\ref{eq:delta}), we can estimate the 
magnification contrast for each lens using 
semi-analytic formulae developed in \citet{inoue-takahashi2012,
takahashi-inoue2014}. We also take into account 
astrometric shifts due to line-of-sight 
structures, which are often overlooked in the literature.
For details, see also \citet{inoue-etal2015}.

\section{Quadruple lens systems}
In this paper, we use 10 QSO-galaxy quadruple lens systems, namely,
B1422+231, B0128+437, MG0414+0534, B1608+656, B0712+472,
B2045+265, H1413+117, PG1115+080, Q2237+0305, RXJ1131-1231
and the SMG-galaxy system SDP.81. We exclude the spiral lens B1933+503,
because it has a very complex structure. Similarly, we also exclude
B1555+375 that has a disc component \citep{hsueh2016}. However, we 
include Q2237+0305 as the bulge component of the spiral lensing 
galaxy seems to dominate over the disk component. 
  In our analysis, we use observed MIR fluxes for MG0414+0534, H1413+117, PG1115+080, Q2237+0305 
and RXJ1131-1231 and radio fluxes averaged over a certain period
for B1422+231, B0128+437, B1608+656, B0712+472, and B2045+265.
For the astrometry of lensed images and the centroids of lensing
galaxies, we use optical or NIR data except for SDP.81 
in order to avoid bias due to complex structures of jets. 
We also use time delay for modeling B1608+656\footnote{We did
not take into account the effect of substructures for estimating the
time delay in our analysis \citep{keeton2009}, 
since the observational errors ($\sim 5 \%$) are too large
to distinguish the perturbative effect.}. 
For SDP.81, we use the processed archival image of
the band 7 continuum taken from the ALMA Science Portal (see
\citet{vlahakis2015} for detail) and for the flux ratios of lensed images, we use fluxes in an aperture radius
0.04 arcsec defined in the source plane. For detail of these 11 lens
systems, see \citet{takahashi-inoue2014, inoue-etal2015,inoue-minezaki2015}.

In table 1, we show the redshifts of the source $z_{\tr{S}}$ and
those of the primary lens $z_{\tr{L}}$, the observed wavelength band of position and flux
of lensed images, the number of lensed images $N$ used for constraining
the lens model. The radius $R_S$ of a MIR continuum source is estimated using dust
reverberation, and $R_S$ of a radio source is estimated from the apparent 
angular size (typically $1\sim 3$ mas in radius) of the lensed very long 
baseline interferometry images. For SDP.81, we adopt the beam size 
in the ALMA band 7 continuum image as the source size, though it may have a
smaller structure. The wavenumber is defined
as $k_S  \equiv 2 \pi/(4 r_S)$, where $r_S$ is the comoving radius of a source.
$k_S$ is used as the upper limit of wavenumber in estimating $\eta$
from line-of-sight structures.


\section{Lens model }

As a fiducial model of a primary lensing galaxy halo, 
we adopt a singular isothermal ellipsoid 
(SIE) \citep{kormann1994}, which can explain
flat rotation curves. There is considerable evidence in
favor of the isothermal profile, and the properties 
of gravitational potential near the Einstein radius 
are insensitive to the radial mass 
distribution \citep{koopmans2009, humphrey2010, barnabe2011, oguri2014}.
We use the fluxes of lensed images, the relative positions of lensed quadruple
images and the centroid of lensing galaxies, and the time delay of lensed
images if available. The contribution from groups, clusters, and large-scale
structures at angular scales larger than the
Einstein angular radius of the primary lens is taken into account as an external
shear (ES). The parameters of the SIE('s) plus ES model 
are the effective Einstein radius $b$ (defined in \citet{kormann1994}), 
the apparent ellipticity $e$ of the lens and its
position angle $\theta_e$, the strength and the direction of the external shear
$(\gamma,\theta_\gamma)$, the lens position $(x_{G},y_{G})$, and the source
position $(x_s,y_s)$. The Hubble constant $h$ is also treated as a model
parameter. The angles $\theta_e$ and $\theta_\gamma$ are measured in
East of North expressed in the observer's coordinates. 
We modeled secondary lenses in MG0414+0534, H1413+117, B1608+656 and
B2045+265 by either an SIS or SIE. 

To find a set of best-fit parameters,
we use a numerical code GRAVLENS \footnote{See
http://redfive.rutgers.edu/$\sim$keeton/gravlens/} developed by Keeton in order to implement
the simultaneous $\chi^2$ fitting of the fluxes, positions, and time delay of lensed
images (if reliable data is available) and the positions of centroid of
lensing galaxies. The best-fitted model parameters are in 
\citet{inoue-takahashi2012, takahashi-inoue2014}.

\section{Method}
In order to measure the magnification perturbation of lensed images, we use a 
statistic $\eta$ defined as
\BE
  \eta \equiv \biggl[\frac{1}{2 N_{\rm pair}} \sum_{{\rm i} \neq {\rm j}} \left[
 \delta^{\tilde{\mu}}_{\rm i} ({\rm minimum}) - \delta^{\tilde{\mu}}_{\rm j} ({\rm saddle})
 \right]^2 \biggr]^{1/2},
\label{eta_def}
\EE
where $\delta^{\tilde{\mu}}({\rm minimum})$ and $\delta^{\tilde{\mu}}({\rm saddle})$ are
magnification contrasts corresponding to the minimum and saddle images and $N_{\rm pair}$
denotes the number of pairs of lensed images. If the correlation
of magnification between pairs of images is 
negligible, then $\eta $ corresponds to the mean 
magnification perturbation per a lensed image. Perturbation by
locally overdense structures would boost $\eta$ as 
the saddle images tend to be demagnified whereas the minimum images are always
magnified provided that the strong lensing effect is negligible.
Note that we need to fix the primary lens model (i.e., 
a best-fitted model without line-of-sight structures or subhaloes) 
in order to calculate $\eta$. In other words, $\eta$ is a model 
dependent statistic though it can be calculated from the observed flux
ratios.

The mean and the standard deviation of $\eta$ can be calculated as follows.
First, we estimate the perturbation  $\varepsilon $ of
the largest angular separation $\theta_{\rm max} $ between
a pair of lensed images X and Y, 
\BE
\varepsilon =|\delta \v{\theta}(\rm{X})-\delta \v{\theta} (\rm{Y})|,
\EE
where $\delta \v{\theta}$ represents the astrometric shift 
of a lensed image at $\v{\theta}$ in the lens plane. 
We then assume that the perturbation satisfies $\varepsilon \le \varepsilon_0$ 
where $\varepsilon_0$ is the observational error for the largest angular
separation. This condition gives an approximated upper limit on the 
contribution of line-of-sight structures or subhaloes assuming that the gravitational 
potential of the primary lens is similar to that of an SIS 
and sufficiently smooth on the scale of the effective
Einstein radius (see appendix). Note, however, that if an astrometric shift caused
by a possible local perturber is parallel to the tangential arc of
a lensed image, though the possibility is small for a point source, the condition is too
strict to some extent\footnote{Some systems with an extended source 
show an Einstein ring in which the magnification is significantly large.
In this case, perturbers in the line of sight to the ring (if present) can cause degeneracy with
the macromodel.}. 

In the case of lensing by line-of-sight structures, we also assume
that small-scale modes with a wavelength larger than  
the mean comoving separation $\langle b \rangle$ between the lens center
and lensed images in the primary lens plane are significantly suppressed. 
The reason is that any modes whose fluctuation scales are larger than
$\langle b \rangle$, which is roughly the size of the comoving effective Einstein
radius, contribute to the smooth components of a 
primary lens, namely, constant convergence and shear. 
Therefore, we consider only modes whose wavenumbers satisfy $k>k_{\rm
lens}$ where $k_{\rm lens} \equiv \pi/(2\langle b \rangle)$. 
Even if the cutoff is applied, sometimes the condition $\varepsilon \le
\varepsilon_0$ cannot be satisfied due to a presence of a possible
secondary lens or other lenses. In this case, we further cutoff the small-scale
fluctuations with a wavenumber $k>k_{\rm lens}$ using the so-called
``constant-shift (CS) cut'' \citep{takahashi-inoue2014}. Roughly
speaking, the CS cut leads to a uniform suppression of astrometric
shift from fluctuations with wavenumbers of $k_{\rm lens}<k<k_{\rm cut}$. 
Using the non-linear power spectrum with the CS cut, we can estimate 
the second moment of $\langle \eta^2 \rangle$. 
From the fitted function of $\eta$ in \citep{takahashi-inoue2014}, 
we can obtain a simple formula $\langle \eta
\rangle \sim 0.85 \langle \eta^2 \rangle^{1/2}$ and the standard
deviation $\delta \eta =0.53 \langle \eta^2 \rangle ^{1/2}$, 
which agrees with the fitted function within a percent level for 
$\langle \eta^2 \rangle^{1/2}<0.2$. Note that the particle masses 
in our $N$-body simulations are $\sim 10^5 h^{-1} \ms$, which can resolve
mass fluctuations at the comoving scale of $\sim 1 \, \tr{kpc}$.

In the case of substructure lensing, we assume that 
subhaloes at the time of infall have an NFW profile. 
Moreover, we also assume that subhaloes that yield astrometric shifts
(at $r \sim r_{-2}$)
larger than $\varepsilon_0$ at the time of infall do not perturb the
lens system. From this constraint, we can 
determine the maximum mass of a perturbing subhalo $m_{200}^{\tr{max}}$. 
These assumptions are reasonable ones as long as the relevant   
subhaloes in the neighbourhood of lensed images 
have a mass profile similar to an SIS. In fact, 
the effect of outer profiles $r>r_{-2}$
of an NFW subhalo can be neglected. The reason is as follows. 
As we shall show in the next section \citep{correa2015}, the 
typical mass scale of subhaloes that can maximally perturb the
positions of lensed images is $\sim 10^9\, \ms$. At the typical infall
redshift $z_{\tr{infall}}\sim 2$, the concentration parameter is $C \sim
7$, which corresponds to $r_{-2} \sim 3\, \tr{kpc}$. Thus, the size
of the region in which the relevant NFW subhaloes have a density profile
with a slope of ${-2}$ is comparable to $\langle b \rangle \sim 5
\, \tr{kpc}$ in our quadruple lens samples. This suggests that these NFW
haloes are kept far from the primary (and the secondary) lens and their effects are limited
to changing modes whose wavelengths are larger than $\langle b
\rangle$. Moreover, for $r>r_{-2}$, the astrometric shifts are smaller than those
at $r \sim r_{-2}$. Note, however, that we also assume that the density
profile of these massive subhaloes at $r\sim r_{-2}$ does not change so much after infall.  
Although the tidal heating causes a concentration of subhalo mass near the center, most of
light rays pass through the outer region of subhaloes where the effect
of tidal heating is much small. Therefore, we expect that the change of inner mass
profile is also not relevant to the astrometric perturbation. 

We also neglect masses below $m_{200}^{\tr{min}}$ for which the perturbation
to the magnification caused by a corresponding SIS (having the same
astrometric shift at $r=r_{-2}$) centred at a source is 
less than $\sim 0.1$ percent due to the finite source 
size $R_S$. In our analysis, for the unperturbed lens component, 
we use $M_{200}$ to make correspondence between a best-fitted SIE 
profile and an NFW profile.

In order to estimate the mass profile of 
subhaloes at the time of infall, we use a semi-analytic model of
NFW haloes (only dynamically relaxed ones) 
in which the concentration parameter $C=C(m_{200},z_{\tr{infall}})$ is given by
\BEA
\log_{10}C&=&a+b \log_{10}(m_{200}/\ms )[1+c (\log_{10}(m_{200}/\ms))^2 ],
\nonumber
\\
a&=&1.7543-0.2766(1+z)+0.02039(1+z)^2,
\nonumber
\\
b&=&0.2753+0.00351(1+z)-0.3038(1+z)^{0.0269},
\nonumber
\\
c&=&-0.01537+0.02102(1+z)^{-0.1475},
\EEA
where $z_{\tr{infall}} \le 4$ \citep{correa2015}. Note that this
relation is also used for calculating $M_{200}$ and $R_{200}$ for the
lensing galaxy haloes.
 
As for the infall redshift, we assume that $0<z_{\tr{infall}}<4$ and
the mean is given by $\langle z_{\tr{infall}} \rangle \sim 2$
\citep{emberson2015}. Since the infall redshift $z_{\tr{infall}}$ of massive subhaloes tends to
be smaller than this value, we take the ambiguity of $\Delta
z_{\tr{infall}}=2 $ into account in estimating the mean of the convergence
perturbation $\delta \kappa$ due to subhaloes.

For estimating the shear perturbation $\delta \gamma$, we assume 
statistical isotropy for the perturbation. Then we have $|\delta
\gamma| \sim |\delta \kappa| $ and we can estimate the magnification 
contrast $\delta \mu$ and $\langle \eta \rangle $ for subhaloes. 
In a similar manner, assuming statistical isotropy,  we can 
estimate the strength of the convergence
perturbation $|\delta \kappa|$ from the observed $\eta$. 
Note that the relation is exact for SISs. Furthermore, we neglect a possible
spatial correlation between perturbers at different positions of lensed
images. The explicit form of calculation is written as follows. 
Let us consider a lens with bright images A (minimum), B (saddle), 
and C (minimum). In terms of observed fluxes
$\mu_{\tr{A}},\mu_{\tr{B}},\mu_{\tr{C}}$ 
 and unperturbed fluxes $\mu_{\tr{A0}},\mu_{\tr{B0}},\mu_{\tr{C0}}$ from a
best-fitted model, the estimator of $\eta $ is given by
\BE
\hat{\eta}^2 = \f{1}{4}\biggl[\biggl( \f{\mu_{\tr{A}} \mu_{\tr{B0}}}
{ \mu_{\tr{A0}}  \mu_{\tr{B}}}-1 \biggr)^2
+\biggl( \f{\mu_{\tr{C}} \mu_{\tr{B0}}}{\mu_{\tr{C0}} \mu_{\tr{B}}}-1
\biggr)^2 \biggr].
\EE
Assuming statistical isotropy, we have
\BE
\hat{\eta}^2 \sim \frac{1}{4}(I_A+2 I_B+I_C) \delta \kappa^2, 
\EE
where 
\BE
I_i\equiv \mu_i^2(4(1-\kappa_i)^2+2 \gamma_i^2),~~for~~ i=\tr{A,B,C},
\label{eq:Ii}
\EE
and $\kappa_i, \gamma_i$ are the non-perturbed 
convergence and the shear at the position of
a lensed $i$-image. Thus, we can estimate $|\delta \kappa|$ from 
lensed images and a best-fitted model.

In the case of
line-of-sight structures, we find that the correlation 
effect is $\sim 10$ percent for lenses with a fold caustic and much
smaller for those with a cusp caustic. Thus, this approximation yields
at most $\sim 10$ percent errors in $|\delta \kappa|$ estimated from $\eta$.   
In the case of subhaloes, we expect that the spatial correlation on
scales of $< 10\, \tr{kpc}$ is negligible after 
infalling to the center of a host halo because of the 
strong tidal force acting on them.

\section{Result}
\begin{table*}
\hspace{-1.5cm}
\begin{minipage}{170mm}
\caption{Quadruple Lens Systems}
\setlength{\tabcolsep}{2pt}
\begin{tabular}{lcccccccccccccr}
\hline
\hline
  lens system  & $z_{\tr{S}}$ & $z_{\tr{L}}$ &position& flux  & $N$ &  $b$($''$) &
 $\langle \kappa \rangle $ 
 &$\log_{10}k_S (h/\tr{Mpc})$ & $R_E$(kpc) & $R_{200}$(kpc) & 
$M_{200}~~~ $ & $m_{200}^{\tr max}~~~ $
& $m_{200}^{\tr min}~~~ $ & reference   \\ 
\hline
  B1422+231 & 3.62 & 0.34  & opt/NIR & radio &
   $3$ & $0.78$ & $0.40$ & $4.8$ & $3.9$ & $320$ & $3.4 \times 10^{12}$
   & $1.1\times 10^8$ & $6.6 \times 10^3$ &(1) (2)
\\
\hline
  B0128+437 & 3.124 & 1.145  & opt/NIR & radio &
   $4$ & $0.24$  & $0.52$ & $4.0$ & $2.1$ & $150$ & $9.2 \times 10^{11}$
   & $1.3\times 10^9$ & $7.1 \times 10^4$ & (1) (3) (4)
\\
\hline
  SDP.81 & 3.042 & 0.2999  & submm & submm &
   $4$ & $1.6$  & $0.50$ & $3.7$ & $7.4$ & $470$ & $1.0 \times 10^{13}$
   & $2.8\times 10^7$ & $4.3 \times 10^5$ & (13)
\\
\hline
  MG0414+0534 & 2.639 & 0.96  & opt/NIR & MIR &
   $4$ & $1.1$ & $0.55$ &  $4.9$ & $9.0$ & $350$ & $9.0 \times 10^{12}$
   & $1.2\times 10^9$ & $3.3 \times 10^3$ & (5) (6) (7)
\\
\hline
  H1413+117 & 2.55 & 1.88($\star$)  & opt/NIR & MIR &
   $4$ & $0.57$ & $0.54$ & $5.5$ & $4.9$ & $270$ & $1.1 \times 10^{13}$
   & $6.2\times 10^9$ & $1.0 \times 10^3$ & (5) (10)
\\
\hline
  PG1115+080 & 1.72 & 0.31  & opt/NIR & MIR &
   $2$ & $1.1$ & $0.52$ & $6.0$ & $5.4$ & $410$ & $7.0 \times 10^{12}$
   & $6.0\times 10^8$ & $1.6 \times 10^2$ & (5) (11)
\\
\hline
  Q2237+0305 & 1.695 & 0.04  & opt/NIR & MIR &
   $4$ & $0.9$ & $0.54$ & $7.1$ & $0.74$ & $370$ & $3.8 \times 10^{12}$
   & $4.1\times 10^8$ & $3.4 \times 10^1$ & (5) (6)
\\
\hline
  B1608+656 & 1.394 & 0.63  & opt/NIR & radio &
   $4$ & $0.91$ & $0.67$ & $3.5$ & $6.4$ & $390$ & $8.6 \times 10^{12}$
   & $5.6\times 10^8$ & $5.8 \times 10^3$ & (2) (8)
\\
\hline
  B0712+472 & 1.339 & 0.406  & opt/NIR & radio &
   $3$ & $0.77$ & $0.50$ & $4.8$ & $4.3$ & $350$ & $4.8 \times 10^{12}$
   & $1.5\times 10^9$ & $7.8 \times 10^3$ & (1) (5)
\\
\hline
  B2045+265 & 1.28 & 0.8673  & opt/NIR & radio &
   $4$ & $0.791$ & $0.39$ & $4.9$ & $8.4$ & $490$ & $2.3 \times 10^{13}$
   & $6.1\times 10^8$ & $9.8 \times 10^3$ & (1) (9)
\\
\hline
  RXJ1131-1231 & 0.658 & 0.295  & opt/NIR & MIR($\star \star$) &
   $3$ & $1.8$ & $0.44$ & $3.9$ & $8.4$ & $640$ & $2.5 \times 10^{13}$
   & $1.0\times 10^9$ & $4.9 \times 10^5$ & (5) (12)
\\
\hline

\label{table1}
\end{tabular}
\\
 Note: The unit of mass for $M_{200}$, $m_{200}^{\tr {max}}$ and 
$m_{200}^{\tr {min}}$ is $h^{-1}\ms$. ($\star$): The lens redshift $z_{\tr{L}}$ is obtained from a best-fit
 model. ($\star \star$): The flux of a line emission [OIII].  References: (1) \citet{koopmans2003} (2) \citet{sluse2012}  (3) \citet{biggs2004} (4) \citet{lagattuta2010}
(5) CASTLES data base:http://www.cfa.harvard.edu/castles
 (6) \citet{minezaki2009}    (7) \citet{macleod2013} (8)
 \citet{fassnacht2002} (9)\citet{mckean2007} (10) \citet{macleod2009}
 (11) \citet{chiba2005} (12) \citet{sugai2007} (13) \citep{inoue-minezaki2015}
\end{minipage}
\end{table*}
\begin{table*}
\begin{minipage}{210mm}
\hspace{-3cm}
\caption{Linear Model Fit of Subsamples (corresponding to red dotted and
 red dashed lines in
 right panels in figure 8)}
\setlength{\tabcolsep}{2pt}
\begin{tabular}{lcccccc}
\hline
\hline
subsample & $\alpha$ & $p$~-~value &  $\beta$ & $p$~-~value
\\
\hline
$z_{\tr{S}}>2$ (with error) & $0.0075\pm 0.0020$ & $0.031$ & $0.0046 \pm 0.0052$
	     & $0.45$
\\
\hline
$z_{\tr{S}}>2$ (without error) & $0.0065\pm 0.0025$ & $0.078$ & $0.0046 \pm 0.0023$
	     & $0.13$
\\
\hline
$z_{\tr{L}}>0.6$ (with error) & $-0.0007\pm 0.0017$ & $0.72$ & $0.0046 \pm 0.0009$
	     & $0.014$
\\
\hline
$z_{\tr{L}}>0.6$ (without error) & $-0.0002\pm 0.0023$ & $0.95$ & $0.0048 \pm 0.0010$
	     & $0.018$
\\
\hline
\end{tabular}
\end{minipage}
\end{table*}

\begin{figure*}
\includegraphics[width=120mm]{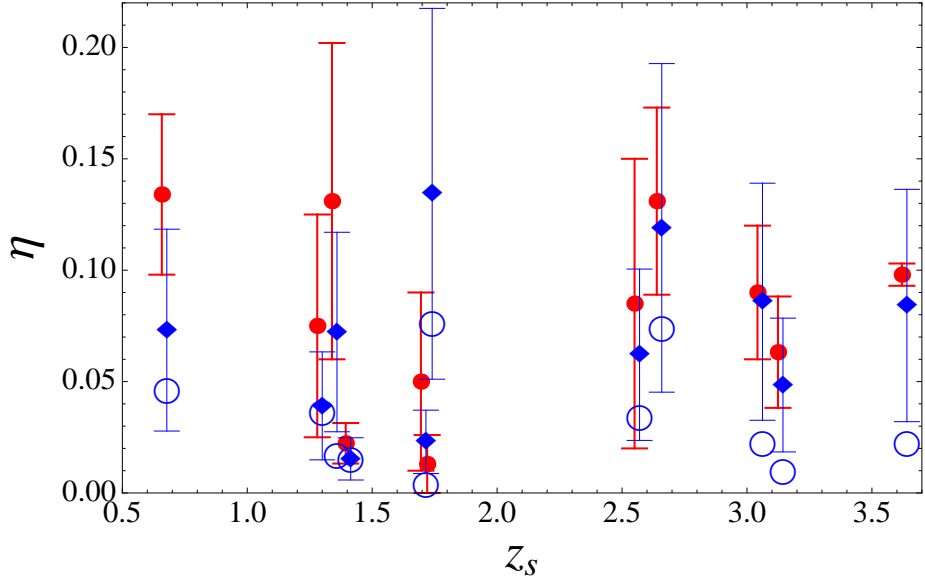}
\caption{Magnification perturbation $\eta$ versus the source redshift
 $z_{\tr{S}}$ for 11 quadruple lens
 samples. Red disks are observed values estimated from the best-fitted
 models with one-sigma observational error bars. Blue diamonds show the 
expected contribution from line-of-sight structures with one-sigma 
error bars. Blue circles show the expected contribution from subhaloes
 in lensing galaxy haloes. The positions of $z_s$ for the 
expected values (blue) are slightly shifted for an illustrative purpose.   }
\label{eta}
\end{figure*}

Firstly, we show the magnification perturbation per a lensed image 
$\eta$ in figure \ref{eta}. The expected contribution $\eta \sim 0.05$ from 
subhaloes is not sufficient to explain the observed 
values $\eta \sim 0.10$ especially for lens systems with a high source redshift $z_{\tr{S}}>3$. 
Although the expected contribution $\eta \sim 0.1$ from line-of-sight structures
is sufficient for explaining observed values, the addition of contribution
from subhaloes improves the fit especially for systems with a low
redshift $z_{\tr{L}}<1.5$ and the overall fit still remains good. 

\begin{figure*}
\includegraphics[width=165mm]{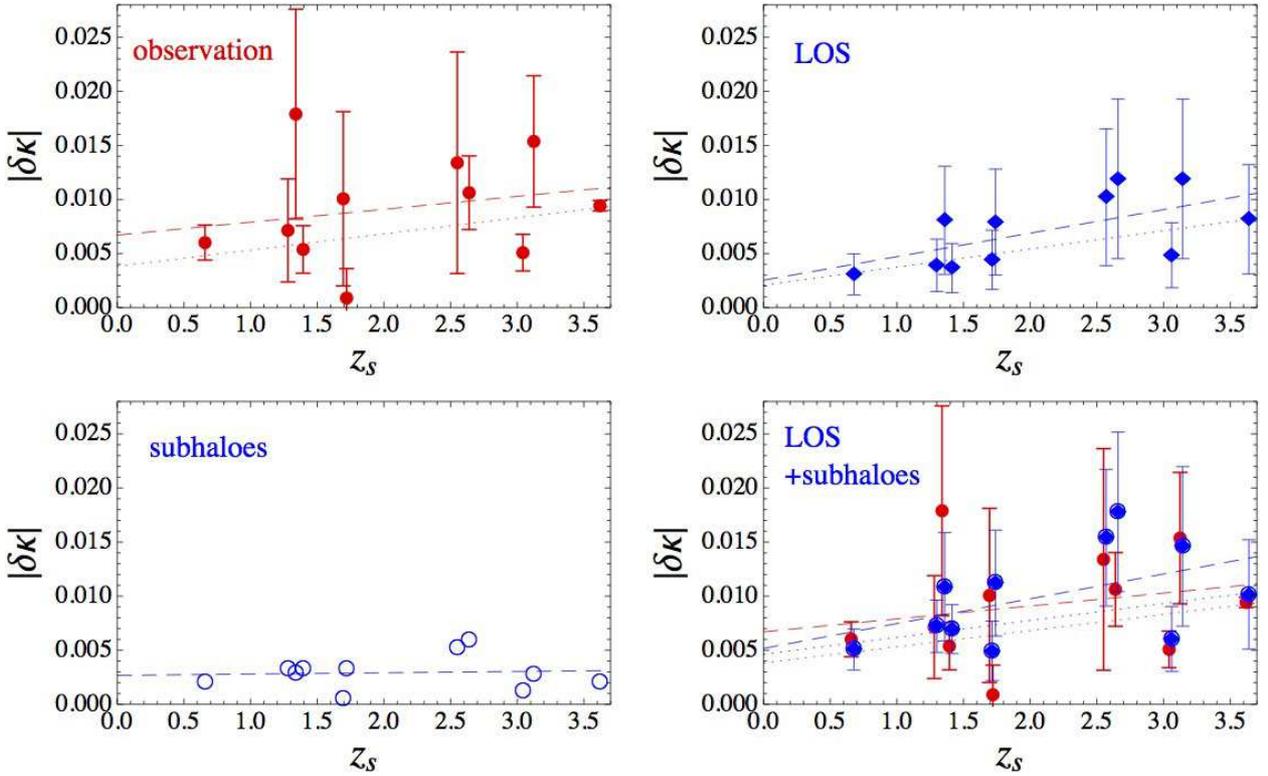}
\caption{The strength of convergence perturbation $\delta \kappa $  versus
 the source redshift $z_{\tr{S}}$ for 11 quadruple lens
 samples. The red disks are observed values estimated from the best-fitted
 models with one-sigma observational error bars. The blue diamonds show the 
expected contribution from line-of-sight structures with one-sigma 
theoretical error bars. The blue circles show the expected contribution from
 subhaloes in a lensing galaxy halo. The errors due to the ambiguity in
the infall time of subhaloes are smaller than the size of the blue circles. 
The blue diamonds embedded in a blue circle show
the expected contribution from line-of-sight structures plus subhaloes 
with one-sigma error bars. Red dotted and red dashed 
lines correspond to linear model fits to the observed values with and 
without taking into account the observational errors, respectively. The positions of $z_s$ for the 
expected values (blue) are slightly shifted for an illustrative purpose. }
\label{deltakappa-obs-zs}
\end{figure*}

Secondly, we show the strength of convergence perturbation $|\delta 
\kappa|$ in $z_{\tr{S}}$ assuming statistical isotropy for perturbations in figure 
\ref{deltakappa-obs-zs}. The observed values (red disks) seem to increase with the 
source redshift $z_{\tr{S}}$. This is consistent with an interpretation that
the dominant contribution comes from the line-of-sight structures (blue diamonds),
because the surface density increases with the comoving length to the
source. On the other hand, the correlation between $\delta \kappa $ from subhaloes 
and $z_{\tr{S}}$ seems to be weak (blue disks). Out of 11 samples, MG0414+0534
shows the largest convergence perturbation $\delta
\kappa=0.0059^{+0.0006}_{-0.0004}$ from subhaloes.
Note that the ambiguity in the infall redshift of subhaloes induces errors of just 5 to 10
percent in $\langle |\delta \kappa| \rangle $, which fall within the size
of circles in figure \ref{deltakappa-obs-zs}. However, the observed
value $|\delta \kappa_{\tr{obs}}|=0.011\pm0.003$ is much larger. For other systems, 
the convergence perturbation due to subhaloes is
much smaller and more difficult to fit to the data. The mean value for
the subhaloes is $\langle \delta \kappa_{\tr{sub}} \rangle =0.003$ whereas 
that for the line-of-sight structures is $\langle |\delta
\kappa_{\tr{los}}| \rangle =0.008$ and that
for the observed values is $\langle |\delta \kappa _{\tr{obs}}| \rangle =0.009$. 
Using the mean
convergence at the positions of lensed images in the best-fitted model, 
the mean of the ratio of a projected dark matter mass fraction in
subhaloes is just $\langle f_{\tr{sub}}\rangle=0.006$ whereas
that for the line-of-sight structures is $\langle f_{\tr{los}}\rangle=0.017$ and that
for the observed values is $\langle f_{\tr{obs}}\rangle=0.019$. 
Thus, it is difficult to 
explain the observed convergence perturbations by contribution from only subhaloes.  

\begin{figure*}
\includegraphics[width=165mm]{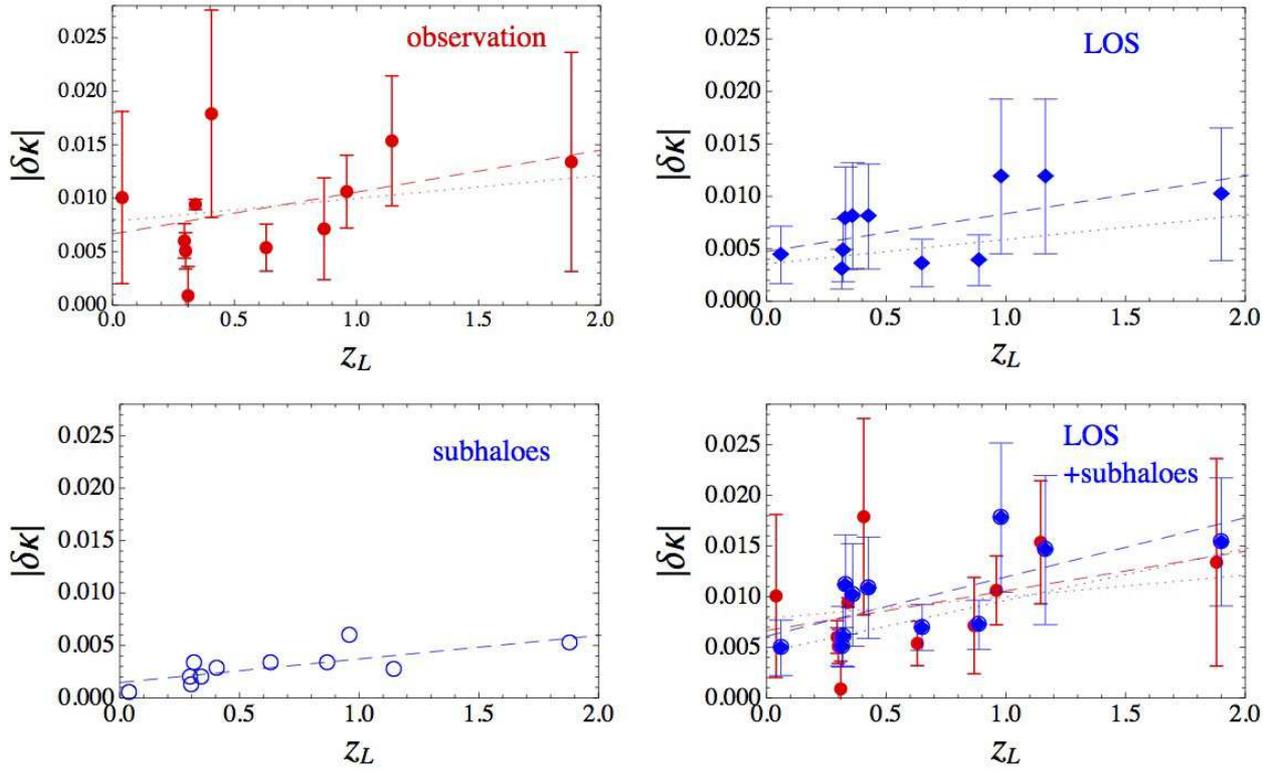}
\caption{The strength of convergence perturbation $\delta \kappa $
 versus the lens redshift $z_{\tr{L}}$ for 11 quadruple lens samples. 
The parameters are the same as in figure \ref{deltakappa-obs-zs}. }
\label{deltakappa-obs-zl}
\end{figure*}

\begin{figure*}
\includegraphics[width=165mm]{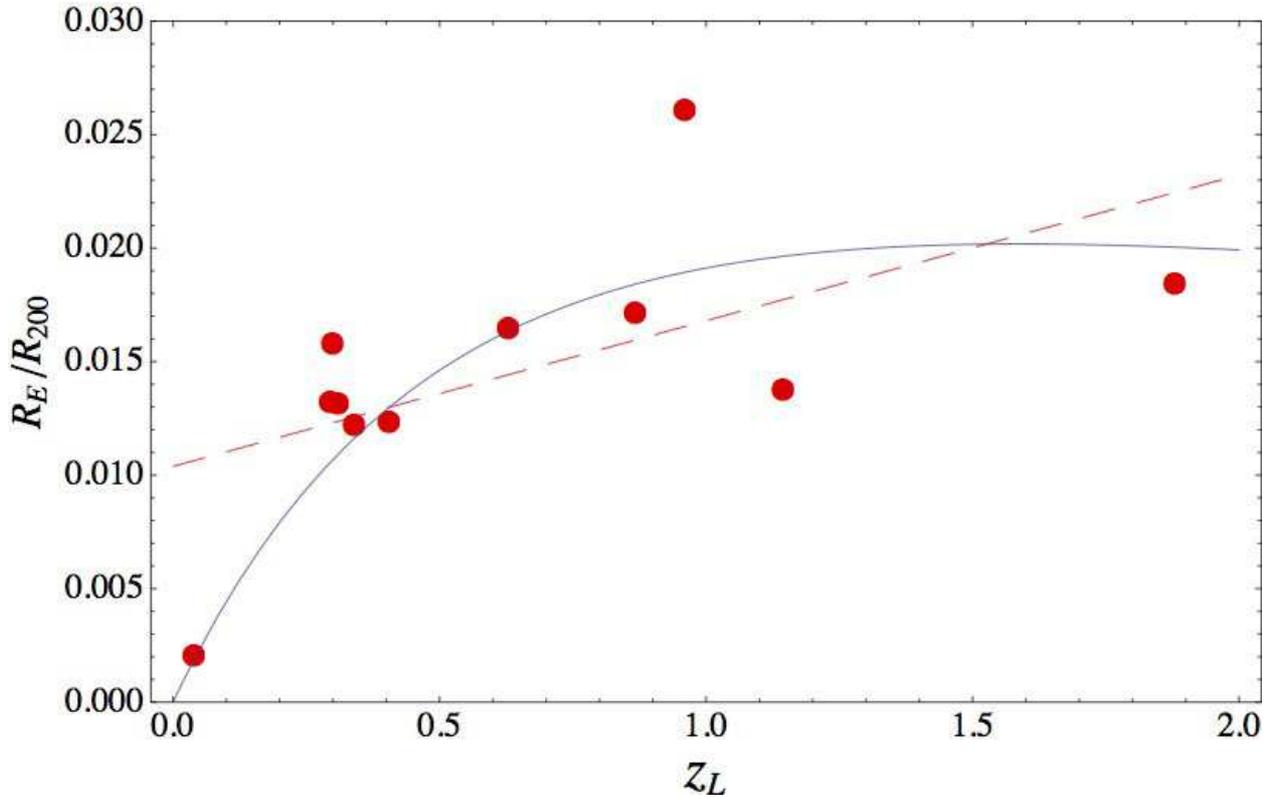}
\caption{The effective Einstein radius $R_E$ in unit of $R_{200}$
of the lensing galaxies (red disks), the linear model fit (red dashed
 line), and the angular diameter distance $D_L$ multiplied by a constant
that is fitted to the data. }
\label{re-r200}
\end{figure*}

\begin{figure*}
\includegraphics[width=165mm]{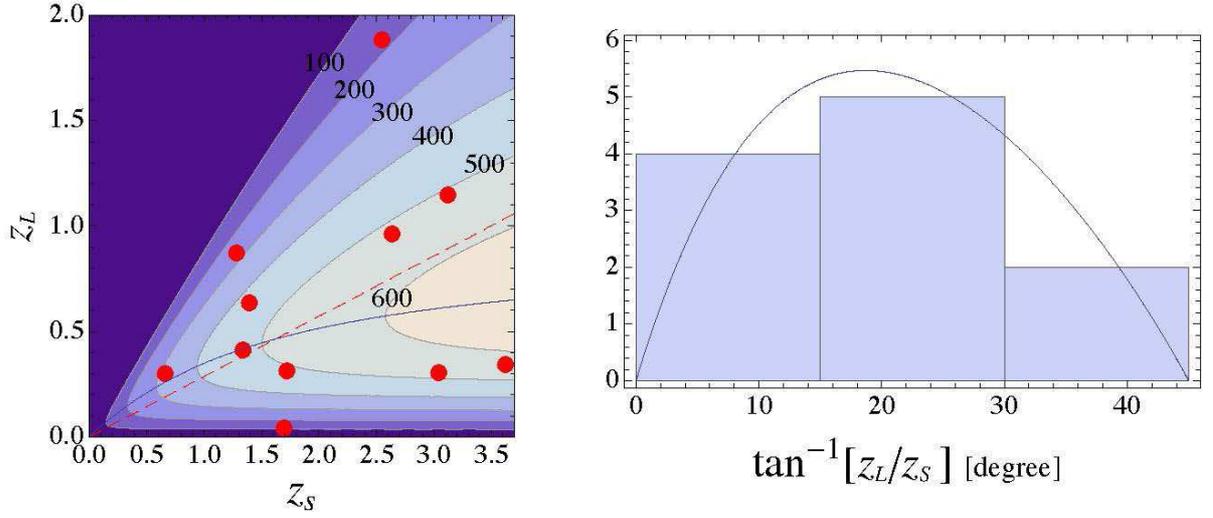}
\caption{The correlation between the source redshift $z_{\tr{S}}$ and the lens
 redshift $z_{\tr{L}}$. Left: the red disks are observed redshifts for 11 samples and 
the red dashed line shows a linear model fit without a constant basis. 
The boundaries of shaded regions represent contours of $L=D_S D_L/D_{LS}$ 
with a decrement of $100 ~h^{-1}\tr{Mpc}$. The blue curve is the ridge
at which $L$ takes its maximum for a given $z_{\tr{S}}$. Right: The histogram
of $\tan^{-1}[z_{\tr{L}}/z_{\tr{S}}]$ of 11 samples and the distribution obtained from
$L$ integrated along $z_{\tr{L}}/z_{\tr{S}}$ for $z_{\tr{S}}<3.6$. }
\label{correlation-zs-zl}
\end{figure*}

Thirdly, we show the strength of convergence perturbation $|\delta 
\kappa|$ in $z_{\tr{L}}$ assuming statistical isotropy for perturbations in figure 
\ref{deltakappa-obs-zl}. It seems that it is consistent with an interpretation that
the dominant contribution comes from subhaloes. This may be due to the fact
that $R_E/R_{200}$ increases with $z_{\tr{L}}$ (figure \ref{re-r200}). The
reason is as follows. Suppose that the effective Einstein angular radius $\theta_E\sim 1''$ is
constant. Then, the effective Einstein radius $R_E=D_L \theta_E $
increases in $z_{\tr{L}}$ until $z_{\tr{L}}\sim 1.5$ (blue curve in figure \ref{re-r200}). Then
$R_E$ begins to decrease slowly due to cosmic expansion. For $z_{\tr{L}}<2.0$,
however, this effect is small. As we have shown in section 2, the ratio
of convergence perturbation of subhaloes $\delta \kappa=\kappa_S$ to the convergence $\kappa_0$
of the lensing galaxy halo increases with $R_E=R_{2D}$. Since 
$\kappa_0\sim 0.5$ (i.e., the lensed images are near the critical curve
in an SIE), we expect that $\delta \kappa$ increases with $R_E$ hence, 
$z_{\tr{L}}$ until $z_{\tr{L}}\sim 1.5$. However, the expected strength is just $|\delta 
\kappa| \lesssim 0.005$ whereas the observed values are $|\delta 
\kappa| \lesssim 0.001$. 
As shown in the upper right panel in figure \ref{deltakappa-obs-zl},
the contribution from line-of-sight structures have a weak dependence 
on $z_{\tr{L}}$. This dependence may come from a weak correlation between
$z_{\tr{S}}$ and $z_{\tr{L}}$. For a given $z_{\tr{S}}$, the possibility of a strong lensing 
is maximum at $z_{\tr{L}}$ where the critical surface density $\Sigma_{\tr{crit}}$,
which is inversely proportional to $L=D_L D_S/D_{LS}$, is the lowest. In
figure \ref{correlation-zs-zl}, 
we plotted redshifts $z_{\tr{S}}$, $z_{\tr{L}}$ of 11 samples with contours of $L$,
which can be regarded as the PDF of strong
lensing. Although the sample is sparse, the histogram of 
angles $\theta=z_{\tr{L}}/z_{\tr{S}}$ has its peak at $\theta=10-20^\circ$, which is
consistent with the result obtained from $L$ integrated along $\theta$. 
Thus, it is reasonable to conclude that the weak $z_{\tr{L}}$ dependence of $\delta
\kappa$ comes from line-of-sight structures as well as subhaloes. 

\begin{figure*}
\includegraphics[width=165mm]{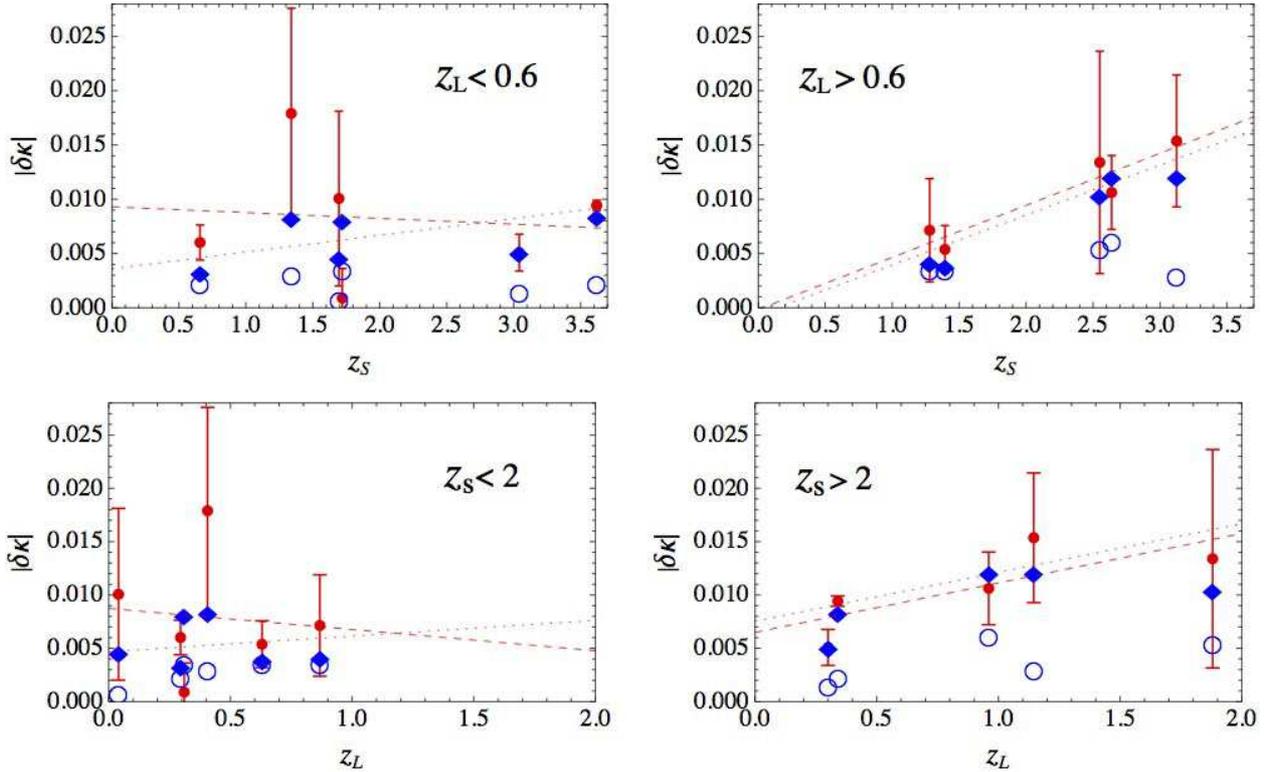}
\caption{The strength of convergence perturbation $\delta \kappa $  versus
 the source redshift $z_{\tr{S}}$ (upper) and the 
lens redshift $z_{\tr{L}}$ (lower) for 11 samples. 
The parameters are the same as in figure \ref{deltakappa-obs-zs}.}
\label{deltakappa-obs-zlzs}
\end{figure*}

Finally, we show an imprint of difference in the redshift dependence of $|\delta
\kappa|$ between $z_{\tr{S}}$ and $z_{\tr{L}}$.  In order to determine the origin 
of anomaly in the flux ratios, we separate the data into a low $z_{\tr{L}}<0.6$ and
a high $z_{\tr{L}}>0.6$ subsamples and a low $z_{\tr{S}} < 2$ and a high $z_{\tr{S}}>2$
subsamples. As one can see in figure \ref{deltakappa-obs-zlzs}, 
a high $z_{\tr{L}}>0.6$ sample shows a strong correlation between $z_{\tr{S}}$ and the
observed $\delta \kappa$ whereas a high $z_{\tr{S}}>2$ sample shows a weak
correlation between $z_{\tr{L}}$ and the observed $\delta \kappa$. The other
subsamples do not show any sign of correlation. In order to
measure the correlation between the redshift and the convergence perturbation, we 
fit the data with a linear function $|\delta \kappa|=\alpha + \beta z_{\tr{L}} (z_{\tr{S}}) $, 
 where $\alpha$ is the base function and $\beta$ is the gradient. As
 shown in table 2, for subsamples with $z_{\tr{L}}>0.6$, a null hypothesis that $\beta = 0$ (no correlation) 
can be rejected at a $98 \sim 99$ percent confidence level. On the other
hand, for subsamples with $z_{\tr{S}}>2$,  the null hypothesis ($\beta = 0$)
cannot be rejected at a $90$ percent confidence level. 
The difference shows that the contribution from line-of-sight structures 
is indeed much larger than that from subhaloes. The apparent correlation
between the observed $|\delta \kappa|$ and $z_{\tr{L}}$ 
is not statistically significant.

\section{Anomalous anomaly}
\begin{figure*}
\includegraphics[width=120mm]{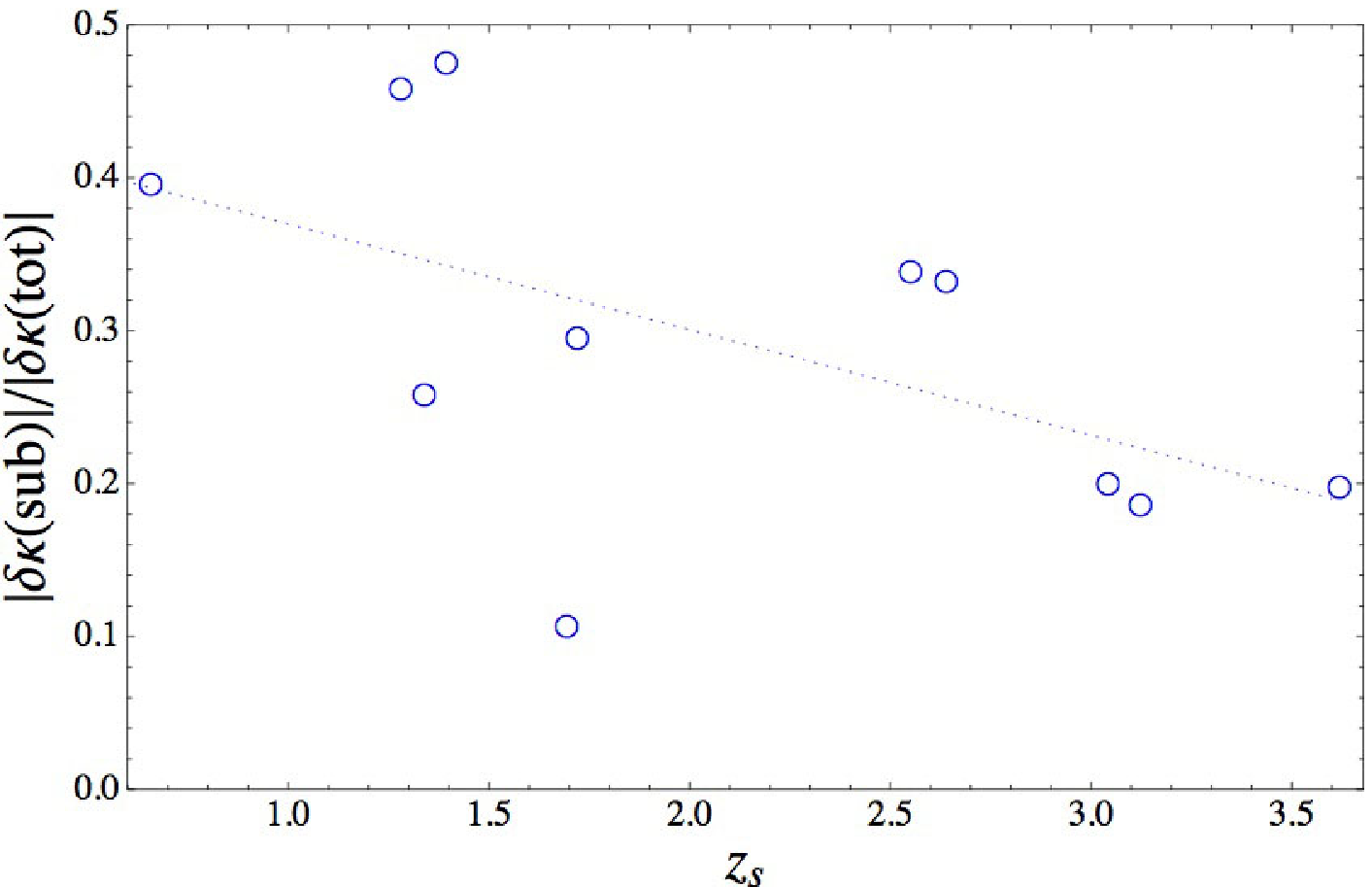}
\caption{The ratio of the strength of 
convergence perturbation from subhaloes to that of the total (subhaloes
 plus line-of-sight structures) shown in blue circles and the linear
 model fit shown in a blue dotted line. }
\label{subhalo-total}
\end{figure*}

If line-of-sight structures are the primary cause of the flux ratio
anomaly, we expect a contribution from positive and negative density 
perturbations almost equally after subtracting the convergence averaged within the
effective Einstein radius \citep{takahashi-inoue2014}. This is due to
the spatial correlation in density fluctuations on scales $\lesssim 1\tr{kpc}$. In the CDM models, the cosmic web structures appear on all the scales
that exceed the free-streaming scale. Therefore, any clumps formed on walls
and filaments should have spatial correlations between them. Moreover, 
intergalactic medium may reside along these structures and
enhance the lensing effect by these clumps due to radiative cooling.
On the other hand, most subhaloes at an effective Einstein radius $R_E$ 
are tidally disrupted. Then, the spatial correlation between subhaloes
are expected to be suppressed, leading to a reduction in the contribution from negative density 
perturbations. Thus, if the contribution from subhaloes is large, we
expect that lensed images of minimum in the arrival time surface tend
to be magnified. However, if the contribution from line-of-sight
structures is large, \ti{there is a $\sim$ 50 percent chance of demagnification in  
minimum images.} If this is the case, the probability of observing such 
an ``anomalous anomaly'' increases with the source redshift $z_{\tr{S}}$. In
order to estimate the probability, we calculate the ratio of the
mean strength of convergence perturbation $\delta \kappa $ from subhaloes to that from
the total (subhaloes and line-of-sight structures). The contribution 
from subhaloes is $\sim 40$ percent of the total at a source redshift
 $z_{\tr{S}}= 0.7$ and decreases monotonically in $z_{\tr{S}}$ to $\sim
20$ percent at $z_{\tr{S}}= 3.6$ (figure \ref{subhalo-total}).  
Assuming that demagnification of a minimum is always caused by  
line-of-sight structures with a 50 percent of chance, 
the probability is $\sim 30$ percent for $z_{\tr{S}}=0.7$ and $\sim 50$
percent for $z_{\tr{S}}=3.6$, the possibility of finding such an ``anomalous anomaly'' 
is expected to be $\sim 30$ percent for $z_{\tr{S}}=0.7$, and $\sim 40$
percent for $z_{\tr{S}}=3.6$. Out of 11 samples, systems with a flux ratio anomaly
at more than 3\,$\sigma $ level are B1442+231, MG0414+0534, and
SDP.81. The ``anomalous anomaly'' is found in B image of SDP.81 \citep{inoue-minezaki2015}.
Therefore, the probability is $\sim 30$ percent and hence 
consistent with the prediction. Although the number
of current sample is too small to verify this
interpretation, a significant increase in the number of quadruple lenses
will lead to discoveries of such a new type of anomaly in the near future. 

\section{Conclusion and Discussion}
In this paper, we have explored the origin of the observed flux ratio anomaly in 
quadruple lensing systems. Firstly, based on a semi-analytical formula, 
we have shown that the surface mass density of subhaloes are nearly constant
at the inner region $R_{2D} < R_{-2}$ of a host galaxy halo where
the light-ray of lensed images pass through. Then we found that 
the expected contribution from 
subhaloes is not sufficient to explain the observed anomalies 
especially for lens systems with a high source redshift $z_{\tr{S}}>3$. 
Although the expected contribution from line-of-sight structures
is already sufficient for explaining observed flux ratios, the addition of contribution
from subhaloes improves the fit especially for systems with a low
redshift $z_{\tr{L}}<1.5$. Secondly, assuming statistical isotropy in
perturbations, the measured convergence perturbation seems to increase with the 
source redshift $z_{\tr{S}}$ as well as the lens redshift $z_{\tr{L}}$. 
 Thirdly, we have shown that the observed correlation between measured
 convergence perturbations with $z_{\tr{S}}$
 is statistically significant at a 98-99 percent confidence level whereas 
that with $z_{\tr{L}}$ is not statistically significant at a 90 percent
confidence level. This result
supports an idea that the dominant contribution comes
from line-of-sight structures rather than subhaloes. We have discussed
a possible origin of correlations based on a semi-analytic formula.
Finally, we have shown that if the contribution from line-of-sight
structures is dominant, 30-40 percent of minimum images that show
anomaly in the relative flux would be systematically demagnified.  
The number of systems showing such an ``anomalous anomaly'' 
first found in SDP.81 \citep{inoue-etal2015} would 
increase with the source redshift $z_{\tr{S}}$.

Our result for substructure lensing is consistent with the previous
result, which was obtained using $R_{\tr{cusp}}$ and $R_{\tr{fold}}$
\citep{xu2015}. Out of 11 lenses, MG0414+0534 showed the largest convergence perturbation
$\delta \kappa=0.0059$ from subhaloes but the value is to small to 
account for the observed flux ratios. We have taken into account
the astrometric shifts due to subhaloes and use a weak+strong lens 
decomposition assuming a smooth potential that is similar to an SIS 
(or SIE) for modeling the unperturbed primary lens. 
Although our approach suffers from inadequate lens modeling,
systematic errors due to the magnification bias and extra lenses do not affect our
constraints. Our obtained value for the mean substructure mass fraction
$\langle f_{\tr{sub}} \rangle  = 0.006$ is consistent with the value $f_{\tr{sub}} =0.0064^{+0.0080}_{-0.0042}$
obtained from SLAC lenses for a substructure mass function slope
$\alpha=0.90(\tilde{\alpha}=1.90)$ \citep{vegetti2014}. Thus, as long as baryonic physics does not
significantly affect the result, it is reasonable to conclude that 
CDM substructures are not the entire reason for the flux anomaly problem.

On the other hand, our result for line-of-sight lensing significantly differs from
the result which was obtained using $R_{\tr{cusp}}$ and $R_{\tr{fold}}$
\citep{xu2012}. The difference may come from the typical source
redshift $z_{\tr{S}}=2$ assumed in \citet{xu2012}. In fact, our result
indicates that for $z_{\tr{S}}<2$, the ratio of the strength of
convergence perturbation from subhaloes to that of the total value
is 30 to 40 percent. For higher source redshifts, however, the ratio becomes much
smaller. Moreover, the two lens systems B2045+265 and B1555+375 used in 
\citep{xu2012} (the latter was excluded in our analysis), which have a relatively large
$R_{\tr{cusp}}=0.4 \sim 0.5$ show another source of anomaly: 
B2045+265 has a secondary lens \citep{mckean2007} and B1555+375 has a edge-on disk
component that can account for the anomaly \citep{hsueh2016}. Excluding
these complex lenses would surely weaken the discrepancy with our
result. Complex lens environments such as secondary lenses and disk
components and neighbouring groups and clusters should be taken into
account properly. 

Our result indicates that a secondary lens can reside
in the intergalactic space and not necessarily a companion of the
primary lens. Furthermore, the previously claimed subhalo with a mass 
of $(1.9 \pm 0.1)\times 10^8 \ms$ discovered
from the residual image of an Einstein ring in B1938+666
\citep{koopmans2005, vegetti2012}
may not be the subhalo of the primary lensing galaxy halo. It can be 
isolated from any haloes or belonging to another distant one.  In order to 
measure the redshift of a secondary or that of a third lens, 
we can use the distortion pattern of an extended source and future submillimetre
observations would be able to detect such small signals \citep{inoue2005a,inoue2005b}.   
With the aid of Atacama Large Millimetre/Submillimetre Array (ALMA), 
detection of other ``anomalous anomaly'' in lensed quadruple SMGs will be
achieved by increasing the number of anomalous targets to $\gtrsim 10$, which
can test our findings. 

 In our analysis, we assumed that primary lenses are described by
SIEs as indicated by the surface brightness of the lensing galaxies.
This assumption may be too simple. We did not take into account slight
smooth distortion of the lens potential, such as low-order multipoles
$m3$ and $m4$ or deviation from isothermal profile. As long as these complexities concern with the fluctuations of the first derivatives on scales comparable to or larger than the Einstein radius of the primary lens, we expect that 
our result would not be changed significantly. This is because 
the anomaly in the flux ratios is caused by perturbation of the second
derivatives of the gravitational potential rather than the first derivatives.  
However, in order to obtain more accurate results, we do need take into
account these complexities as well. These issues will be addressed in 
our future work.   

\section{Acknowledgments}
The author thanks Masashi Chiba and Takeo Minezaki for valuable discussion and useful comments. 
This work is supported in part by JSPS Grant-in-Aid for
Scientific Research (B) (No. 25287062) ``Probing the origin
of primordial minihalos via gravitational lensing phenomena''. This
paper makes use of the following ALMA data:
ADS/JAO.ALMA\#2011.0.00016.SV. ALMA is a partnership of ESO (representing
its member states), NSF (USA) and NINS (Japan), together with NRC
(Canada), NSC and ASIAA (Taiwan), and KASI (Republic of Korea), in
cooperation with the Republic of Chile. The Joint ALMA Observatory is
operated by ESO, AUI/NRAO and NAOJ.
\appendix

\section{Uniqueness in Decomposition of Strong+weak Lens}
In what follows, we probe uniqueness in decomposition
of a strong plus weak lens that creates a quadruple image.
This proof ensures that if deflection angle caused by perturbers 
is too large, reparametrisation of the best-fit
SIE model cannot recover the fit to all the image positions.
We assume that the gravitational potential $\Psi$ of a primary lens 
is sufficiently smooth and similar to that of an SIS and the lens is 
successfully fitted. We show
that reparametrisation of dimensionless model parameters $\largec$ and the source
angular position $\y$ of the best-fitted model with
$\Psi(\largec)$ does not recover the original fit in the positions 
if the $j$-th lensed image at an angular position of $\x_j$ is 
significantly perturbed to $\x_j+\delta \x_j$, where  $|\delta \x_j|$
far exceeds the observational error $\varepsilon$ except for a special case 
in which $\delta \x_j$ is parallel to the tangential arc. 
For brevity without loss of generality, we set the effective Einstein radius of the best-fitted 
model to $1$ and all the $1 \sigma$ errors of observed positions 
of lensed images to a constant $\varepsilon \ll 1$. 

The unperturbed lens equation for the $i$-th image $(i=1,2,3,4)$ is
\BE
\y = \x_i - \ALP (\x_i; \largec),
\label{eq:lens}
\EE
where $\ALP (\x_i; \largec)$ denotes the deflection angle obtained from $\Psi(\largec)$.
Suppose that the $j$-th lensed image placed at the neighbourhood of 
a critical curve in the unperturbed lens is affected by a local perturber so that the position
of the $j$-th lensed image is shifted from $\x_j$ to $\x'_j=\x_j+\delta
\x_j$ where $|\delta \theta_j|=\eta \gg \varepsilon$ whereas other lensed
images at $\x_i, (i \neq j)$ remain the observed positions within $\varepsilon$. Then, we need to adjust
the position $\x'_j$ back to $\x_j$ by changing the lens parameter
$\largec$ to $\largec '=\largec+\delta \largec$ and the source position 
$\y$ to $\y'=\y+\delta \y$ keeping the positions of other lensed images
within $\varepsilon$.  From the perturbed lens equation and
the unperturbed lens equation $(\ref{eq:lens})$ , we have 
\BE
 \delta \y \approx -\f{\del \y}{\del \x}\biggr|_{\x_j}\delta \x_j - \f{\del \ALP_j}{\del C} \delta \largec
\label{eq:A2}
\EE
where $\ALP_j=\ALP(\x_j;\largec)$. Using an orthogonal transformation,
the inverse magnification matrix $\del \y/\del \x $ can be
diagonalized as 
\BE
\f{\del \y}{\del \x}=
\begin{pmatrix}
1-\kappa-\gamma & 0 \\
0 &  1-\kappa+\gamma \\
\end{pmatrix},
\EE
where $\kappa$ and $\gamma$ are the convergence and shear, respectively.
Assuming that the lensed images are significantly magnified and
$\Psi$ is similar to that of an SIS, we have $\kappa=\gamma\sim 0.5$. 
Then, we have
\BE
\f{\del \y}{\del \x} \sim \begin{pmatrix}
0 & 0 \\
0 & 1 \\
\end{pmatrix},
\label{eq:A4}
\EE 
and thus the order of $(\del \y/\del \x) \delta \x_j $ is $\eta$ 
except for the case in which $\delta \x_j \sim (\delta x_j, 0)$, i.e.,
parallel to the tangential arc. On the other hand, 
$O[(\del \y/\del \x) \delta \x_i]=\varepsilon,  (i \neq j) $ or
smaller ($O$ denotes order). Since 
\BE
 \f{\del \y}{\del \x}\biggr|_{\x_i}\delta \x_i +
 \f{\del \ALP_i}{\del C} \delta \largec
=\f{\del \y}{\del \x}\biggr|_{\x_k}\delta \x_k +
 \f{\del \ALP_k}{\del C} \delta \largec,
\EE
for $i\ne j$ and $k \ne j$, and  
\BE
O \biggl[\biggl| \f{\del \ALP_i}{\del \largec}\delta \largec
\biggr|\biggr]=O\biggl[\f{\alpha_i}{ C}\delta C
\biggr]=O\biggl[\f{\delta C}{C} \biggr],
\EE 
because of smoothness of $\Psi$, we have $O[\delta C/C ] \le \varepsilon$
and $O[|\delta \y|] \le \varepsilon$. However, equation (\ref{eq:A2})
and (\ref{eq:A4}) give $O[|\delta \y|] = \eta$.  Thus, we cannot find
any reparametrisation except for the case in which $\delta \x_j$ is
parallel to the tangential arc. In other words, a massive local perturber 
that causes a large astrometric shift makes
the unperturbed model significantly differ from a smooth SIS.

\bibliographystyle{mnras}
\bibliography{weak-lensing-by-los2016}

\end{document}